\documentstyle[12pt]{article}
\begin{document}

\title{All static spherically symmetric anisotropic solutions of Einstein's equations}

\author{ L. Herrera$^1$\thanks{e-mail:
laherrera@cantv.net.ve}, J. Ospino$^{2}$\thanks {e-mail:
jhozcrae@usal.es} and  A Di Prisco$^1$\thanks{e-mail:
adiprisc@fisica.ciens.ucv.ve} \\
\small{$^1$Escuela de F\'{\i}sica, Facultad de Ciencias,} \\
\small{Universidad Central de Venezuela, Caracas, Venezuela.}\\
\small{$^2$Area de F\'\i
sica Te\'orica. Facultad de Ciencias,Universidad de Salamanca,}\\
\small{Salamanca, Spain.}\\
}

\maketitle

\begin{abstract}
An algorithm  recently presented by Lake  to obtain all static spherically symmetric perfect fluid solutions, is extended to the case of locally anisotropic fluids (principal stresses unequal). As expected, the new formalism requires the knowledge of two functions (instead of one) to generate all possible solutions. To illustrate the method some known cases are recovered.
\end{abstract}

\maketitle

\newpage

\section{Introduction}
As is well known, static spherically symmetric perfect fluid distributions in general relativity, are described by a system of  three independent Einstein equations for four variables (two metric functions, the energy density and the isotropic  pressure).  Thus, aditional information in the form of an equation of state or an heuristic assumption  involving metric and/or physical variables has to be provided in order  to integrate the system. This situation suggests the  possibility of  obtaining any possible solution, giving a single  generating function. A formalism to obtain solutions in  this way has been recently presented by Lake \cite{KLake} (see also \cite{viser}).

The purpose of this work is to extend the above mentioned formalism to the case of locally anisotropic fluids. 

The motivation for doing so  is provided by the fact that  the assumption of local anisotropy of pressure,
which seems to be very reasonable for describing the matter distribution
under a variety of circumstances, has been proved to be very
useful in the study of relativistic compact objects (see
\cite{PR}- \cite{N}  and references therein).

In the next section we shall present the  general equations and the formalism to obtain the solutions, then  we shall apply  the method to analyze some specific cases.

\section{The Einstein equations for static locally anisotropic fluids}

In curvature coordinates the line element reads
\begin{equation}
ds^2=-e^{\nu (r)} dt^2+e^{\lambda (r)}
dr^2+r^2d\theta^2+r^2sin^2\theta d\phi^2  \label{metric}
\end{equation}

 \noindent which has to satisfy the Einstein equations. For  a locally anisotropic fluid they are
\begin{equation}
8\pi \rho=\frac{1}{r^2}-e^{-\lambda}
\left(\frac{1}{r^2}-\frac{\lambda'}{r} \right), \label{fieq00}
\end{equation}

\begin{equation}
8\pi  P_r =-\frac{1}{r^2} +e^{-\lambda}
\left(\frac{1}{r^2}+\frac{\nu'}{r}\right), \label{fieq11}
\end{equation}

\begin{eqnarray}
8\pi P_\bot = \frac{e^{-\lambda}}{4} \left(2\nu''+\nu'^2 -
\lambda'\nu' + 2\frac{\nu' - \lambda'}{r}\right), \label{fieq2233}
\end{eqnarray}
where primes denote derivative with respect to $r$, and $\rho, P_r$ and $P_\bot$ are the proper  energy density,  radial pressure and  tangential pressure respectively.
\subsection{The algorithm}
\noindent From (\ref{fieq11}) and  (\ref{fieq2233}) it follows:

\begin{equation}
8\pi(P_r-P_\bot)=e^{-\lambda}(-\frac{\nu^{\prime
\prime}}{2}-(\frac{\nu^{\prime}}{2})^2+\frac{\nu^{\prime}}{2r}+\frac{1}{r^2})+e^{-\lambda}\frac{\lambda^{\prime}}{2}(
\frac{\nu^{\prime}}{2}+\frac{1}{r})-\frac{1}{r^2}. \label{anis}
\end{equation}

\noindent Then, introducing the variables 
\begin{equation}
e^{\nu (r)}=e^{\int (2z(r)-2/r)dr} 
\label{v1}
\end{equation}

and 
\begin{equation}
e^{-\lambda}=y(r)
\label{v2}
\end{equation}
and feeding back into  (\ref{anis}) we get:

\begin{equation}
y^{\prime}+y[\frac{2z^\prime}{z}+2z-\frac{6}{r}+\frac{4}{r^2
z}]=-\frac{2}{z}(\frac{1}{r^2}+\Pi(r)), \label{eq1}
\end{equation}

\noindent with $\Pi(r)=8\pi (P_r-P_\bot).$

\noindent Integrating (\ref{eq1}) we obtain for  $\lambda$:

\begin{equation}
e^{\lambda (r)}=\frac{z^2(r) e^{\int(\frac{4}{r^2 z(r)}+2z(r))dr}}
{r^6(-2\int\frac{z(r)(1+\Pi (r)r^2 ) e^{\int(\frac{4}{r^2
z(r)}+2z(r))dr}}{r^8}dr+C)}.\label{lambda}
\end{equation}
where $C$ is a constant of integration.
\noindent Then, using   (\ref{v1}) and (\ref{lambda}) in (\ref{metric})
we get:
\begin{equation}
ds^2=-e^{\int (2z(r)-2/r)dr}dt^2+\frac{z^2(r) e^{\int(\frac{4}{r^2
z(r)}+2z(r))dr}} {r^6(-2\int\frac{z(r)(1+\Pi (r)r^2)
e^{\int(\frac{4}{r^2
z(r)}+2z(r))dr}}{r^8}dr+C)}dr^2+r^2d\theta^2+r^2sin^2\theta
d\phi^2. \label{metric2}
\end{equation}
Thus any solution describing a static  anisotropic fluid distribution is fully determined by means of the two generating functions $\Pi$ and $z$.

It will be convenient to express the physical variables in terms of metric and  generating functions, in order to impose conditions leading to physically meaningful solutions.
Thus we have:
\begin{equation}
4\pi P_r=\frac{z(r-2m)+m/r-1}{r^2}\label{Pr}
\end{equation}
\noindent 

 \begin{equation}
4\pi \rho =\frac{m^{\prime}}{r^2}\label{rho}
\end{equation}
and 
\begin{equation}
4\pi P_\bot=(1-\frac{2m}{r})(z^{\prime}+z^2-\frac{z}{r}+\frac{1}{r^2})+z(\frac{m}{r^2}-\frac{m^{\prime}}{r})
\label{Pbot}
\end{equation}
where the mass function $m(r)$ is defined as usual by 
\begin{equation}
e^{-\lambda}=1-\frac{2m(r)}{r}
\label{m}
\end{equation}

Physically meaningful solutions must be regular at the origin, and should satisfy the conditions $\rho>0$, $\rho>P_r, P_\bot$. If stability is required then $\rho$ and $P_r$ must be monotonically decreasing functions of $r$.

To avoid singular behaviour of physical variables on the boundary of the source ($\Sigma$), solutions should also satisfy the Darmois conditions on the boundary .  Implying $(P_r)_\Sigma=0$ and 
\begin{equation}
e^{\nu_\Sigma}=e^{-\lambda_\Sigma}=1-\frac{2M}{r_\Sigma}
\label{ens}
\end{equation}
 with $m_\Sigma=M$, and $r_\Sigma$ denotes the radius of the fluid distribution.

\subsection{The locally isotropic case}

If we impose the isotropic condition on pressure
\begin{equation}
 \Pi=8\pi(P_r-P_\bot)=0
\end{equation}
in  (\ref{metric2}) we obatin:

\begin{equation}
ds^2=-e^{\int (2z(r)-2/r)dr}dt^2+\frac{z^2(r) e^{\int(\frac{4}{r^2
z(r)}+2z(r))dr}} {r^6(-2\int\frac{z(r)e^{\int(\frac{4}{r^2
z(r)}+2z(r))dr}}{r^8}dr+C)}dr^2+r^2d\theta^2+r^2sin^2\theta
d\phi^2 \label{metric3}
\end{equation}

\noindent which is  the same result obtained in \cite {KLake},with  $z(r)=\Phi(r) ^{\prime}+\frac{1}{r}$.

\section{Some examples}
We shall next apply the algorithm to reproduce some known situations.
\subsection{Conformally flat anisotropic fluids}

Instead of giving two generating functions, we may provide one generating function and an additional {\it{ansatz}}. Thus for example, in the spherically symmetric  case we know that there is only one independent component of the Weyl tensor.  Therefore the conformally flat condition reduces to a single equation which reads\begin{equation}
\frac{\nu ^{\prime
\prime}}{2}+(\frac{\nu ^{\prime}}{2})^2-\frac{\nu^{\prime}\lambda
^{\prime}}{4}-\frac{\nu^{\prime}-\lambda^{\prime}}{2r}+\frac{1-e^{\lambda}}{r^2}=0.\label
{E}
\end{equation}

\noindent Equation  (\ref{E}) has been integrated in \cite{j1},
giving:
\begin{equation}
e^{\frac{\nu}{2}}=c\,r \,cosh(\int
\frac{e^{\frac{\lambda}{2}}}{r}dr),\label{nulambda}
\end{equation}
\noindent which, in term of $z$ becomes
\begin{equation}
z=\frac{2}{r}+\frac{e^{\frac{\lambda}{2}}}{r}\, tanh(\int
\frac{e^{\frac{\lambda}{2}}}{r}dr).\label{zlambda}
\end{equation}
\noindent On the other hand from  (\ref{fieq2233}) and
(\ref{E}), it follows:
\begin{equation}
\Pi=r(\frac{1-e^{-\lambda}}{r^2})^\prime.
\label{Pilambda}
\end{equation}
\noindent Thus the system is completely determined  (in this case)  provided a single generating function $z$ is known.

\subsection{Bowers-- Liang solution}
\noindent This solution corresponds to an anisotropic fluid with an homogeneous energy density distribution $\rho=\rho_0=constant$ \cite{Bowersliang}, and is given by: 
\begin{equation}
e^\nu = \left[\frac{3 \left(1 - 2M/r_\Sigma\right)^{h/2} - 
\left(1 - 2m/r\right)^{h/2}}{2}\right]^{2/h}
\label{esn}
\end{equation}

\begin{equation}
m(r) = \frac{4 \pi}{3} r^3 \rho_0 \qquad ; \qquad 
M = \frac{4 \pi}{3} r_\Sigma ^3 \rho_0
\label{mM}
\end{equation}

The two generating functions for this metric are:
\begin{equation}
z=\frac{\frac{2m}{r^2}(1-\frac{2m}{r})^{\frac{h}{2}-1}}{3(1-\frac{2M}{r_\Sigma})^{\frac{h}{2}}-(1-\frac{2m}{r})^{\frac{h}{2}}}+\frac{1}{r}
\label{ok}
\end{equation}

and
\begin{equation}
\Pi=-6C\frac{(z-\frac{1}{r})^2(1-\frac{2M}{r_\Sigma})^{\frac{h}{2}}}{(1-\frac{2m}{r})^{\frac{h}{2}-1}}
\label{k}
\end{equation}
with $h=1-2 C=Constant$. The case $h=1$ reproduces  the well known Schwarzschild interior solution, whereas the case $h=0$  describes the Florides solution \cite{florides}.

\subsection{Anisotropic solutions with a non--local equation of state}
\noindent An interesting family of solutions may be found from the assumption that the energy density and the radial pressure are related by a non--local equation of state  of the form  \cite{HHLN1}
\begin{equation}
P_r(r)=\rho(r)-\frac{2}{r^3}\int_{0}^{r}\tilde r^2\rho(\tilde
r)d\tilde r+\frac{C}{2\pi r^3}\label{NLES}
\end{equation}
\noindent or, using(\ref{rho})
\begin{equation}
P_r(r)=\frac{m^\prime}{4\pi r^2}-\frac{m}{2\pi r^3}+\frac{C}{2\pi
r^3} \label{NLES1},
\end{equation}

\noindent From  (\ref{Pr}) and  (\ref{NLES1}) it follows that these solutions are defined by the generating function  $z$ of the form:
\begin{equation}
z=\frac{rm^\prime-3m+2C+r}{r(r-2m)}. \label{zNLES}
\end{equation}

\section*{Acknowledgments.}
JO acknowledges financial support from the Junta de Castilla y Leon (Spain)  under grant SA010C05. LH wishes to thank Kayll Lake for pointing out (after the completion of this manuscript)  a previous paper \cite{Lake2} where the anisotropic case is considered.

\end{document}